\documentclass[preprint,noshowpacs,aps,amsfonts,showpacs,superscriptaddress,floatfix]{revtex4}
\usepackage{graphicx} 
\usepackage{amsmath}
\usepackage{amssymb}

\begin{document}

\title{Relaxation dynamics of conserved quantities in a weakly non-integrable one-dimensional Bose gas}

\author{Giuseppe Brandino}
\affiliation{Institute for Theoretical Physics, Science Park 904, University of Amsterdam, 1090 GL
Amsterdam, The Netherlands}
\author{Jean-S\'ebastien Caux}
\affiliation{Institute for Theoretical Physics, Science Park 904, University of Amsterdam, 1090 GL
Amsterdam, The Netherlands}
\author{Robert Konik}
\affiliation{CMPMS Dept., Brookhaven National Laboratory, Upton, NY 11973-5000, USA}

\date{\today}

\begin{abstract}
In this work we report preliminary results on 
the relaxational dynamics of one dimensional Bose gases, as described by the Lieb-Liniger model, 
upon release from a parabolic trap.  We explore the effects of integrability and integrability
breaking upon these dynamics by placing the gas post-release in an integrability breaking
one-body cosine potential of variable amplitude.  By studying
the post-quench evolution of the conserved charges that would exist in the purely integrable limit,
we begin to quantify the effects of the weak
breaking of integrability on the long time thermalization of the gas.
\end{abstract}

\pacs{03.75.Kk, 03.75.Hh, 05.30.Jp, 67.10.-d}
\keywords{Lieb-Liniger model, renormalization, quenches}

\maketitle

\section{Introduction}

In their seminal experiment on one dimensional interacting cold atomic gases, 
Kinoshita et al. \cite{weiss}
argued for the possibility that the relaxational dynamics of such gases possessed memory of
the gases' initial condition.  Specifically, they observed that the momentum
distribution of the gas did not rapidly
evolve to a thermal equilibrium state, despite the presence of interactions between the gas' atoms.  
To explain this behavior, they conjectured that the gas possessed
non-trivial conserved integrals of motion (beyond the energy of the gas), and 
that these integrals of motion were controlling
the long time dynamics of the gas.  These non-trivial integrals of motion should exist, they argued,
because the underlying theoretical description of the gas, the Lieb-Liniger model, is known to be exactly
solvable and has an infinite set of such integrals \cite{ll,korepin}.   
In subsequent work Rigol et al. \cite{rigol} sharpened this conjecture
by arguing that while the gas did relax to a state governed by a thermodynamic ensemble, 
this ensemble was not the canonical (or microcanonical), but an ensemble aware of these additional
integrals of motion, an ensemble they dubbed the generalized Gibbs ensemble.

In subsequent work this ensemble has been shown to govern the dynamics of a number of systems characterized
by sets of non-trivial conserved quantities, both non-interacting \cite{barthel,caza,caza1,essler,rossini}
and interacting \cite{fior_muss,NRG4}.
However less studied has been the question of thermalization when the system has a set of 
weakly broken integrals of motion \cite{yur_ols,canovi,kollar}.  Does the weak breaking of the integrals of motion always lead to eventual
thermalization of the gas as governed by the canonical ensemble?  Is there a time scale of
integrability breaking, $\tau_{IB}$, for which for times $t < \tau_{IB}$ the dynamics appear integrable
while for times $t> \tau_{IB}$, the dynamics are governed by the standard thermodynamic ensembles?
Or is there a smooth crossover in thermalization as suggested in \cite{yur_ols,kollar,canovi}, where physical
quantities interpolate between their values under the generalized Gibbs ensemble and their values in the
canonical ensemble?

It is with these questions in mind that we study the following quantum quench problem.  We begin by 
considering a one dimensional atomic Bose gas 
of $N$ particles in a system of length $L$ in the presence of
a one-body parabolic trap, $V_{\rm para}(x)=m\omega^2x^2/2$.  
We describe the gas with the Hamiltonian,
\begin{eqnarray}
H &=& H_{LL} + \sum_{i=1}^NV_{\rm para}(x_i);\cr\cr
H_{LL} &=& -\frac{\hbar^2}{2m}\sum^N_{j=1}\frac{\partial^2}{\partial x_j^2}+2c\sum_{\langle i,j\rangle}\delta(x_i-x_j),
\end{eqnarray}
where we work in units of $2m=\hbar=1$.  $H_{LL}$ is the well known Lieb-Liniger model \cite{ll}
with interaction strength $c$.    
We prepare the system in the ground state of this Hamiltonian, $H$.
At $t=0$ we remove the parabolic trap.  In the absence of the parabolic trap, the Hamiltonian 
(now just the Lieb-Liniger model itself)
is integrable, and we expect the subsequent dynamics of the gas to be governed by the non-trivial
conserved charges in the system.  If instead of simply removing the trap, we replace it with a different
one-body potential, $V_{\rm cos}(x)= A\cos (\omega x)$, 
we break integrability, so changing the nature of the post-quench dynamics.
By varying the amplitude $A$ of this potential, we control the amount of integrability breaking in the
system and its concomitant effects on the dynamics.  This quantum quench is illustrated in Fig. 1.

To study these dynamics, we employ a combination of Bethe ansatz solvability and a numerical renormalization
group.  The Lieb-Liniger model is a model that can be solved
with the Bethe ansatz, both determining its spectra \cite{ll} and its matrix elements \cite{1989_Slavnov_TMP_79_82}.
Because we can compute matrix elements, we can compute correlations functions of this model using 
their Lehmann representations \cite{caux,cc}.
This however is computationally intensive.  In order to accomplish this task we employ an optimized set of routines
known as ABACUS \cite{caux}.

\begin{figure}[t]
\centering
\includegraphics[scale=0.6]{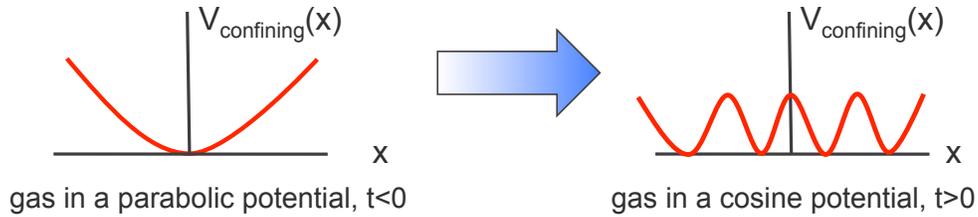}
\vskip -2.5in
\caption{Sketch of the quench protocol.  For $t<0$, we prepare the gas in its ground
state in a parabolic potential.  At $t=0$, we remove the parabolic potential and replace it 
with a small cosine potential.}
\end{figure}

Because we are interested in perturbing the gas by introducing integrability breaking one-body potentials, 
both pre- and post-quench, the solvability of the Lieb-Liniger model is insufficient for the task at hand.
Instead to study such deformations, 
we employ a numerical renormalization group (NRG) \cite{NRG,NRG1,NRG2,NRG3,NRG4} able
to study perturbations of integrable and conformal continuum field theories.  This 
approach, as it is an extension of a methodology known as the truncated conformal spectrum
approach \cite{YZ,YZ1}, has been primarily used to study perturbations of relativistic field theories \cite{NRG,NRG1,NRG2,NRG3}, but
has recently been applied to the Lieb-Liniger model perturbed by a one-body potential \cite{NRG4}, the problem
at hand.  The NRG uses the eigenstates of the Lieb-Liniger model as a computational basis.  Because this
basis accounts for the interactions of the Bose gas particles with one another, this numerical method builds
in the strong correlations present in the problem right at the start.

The paper is organized as follows.  In Section 2 we demonstrate that we can compute the equilibrium properties
of the gas in the one-body potentials.  While we have shown in Ref. \cite{NRG4} that we can accurately describe
the ground state and first few excited states in such a potential, here we show that we can obtain with reasonably accuracy a wide
range of the spectrum.  This will be important for the determination of the post-quench time evolution of the
Lieb-Liniger conserved charges.
In Section 3, we consider the evolution of the expectation values of these charges.
This will allow us to begin to understand
the consequences of integrability breaking.  In Section 4, we discuss these results briefly and examine
possible further directions for this work.


\section{Equilibrium properties of the gas in the one-body potentials}

\begin{figure}[t]
\centering
\includegraphics[scale=0.2]{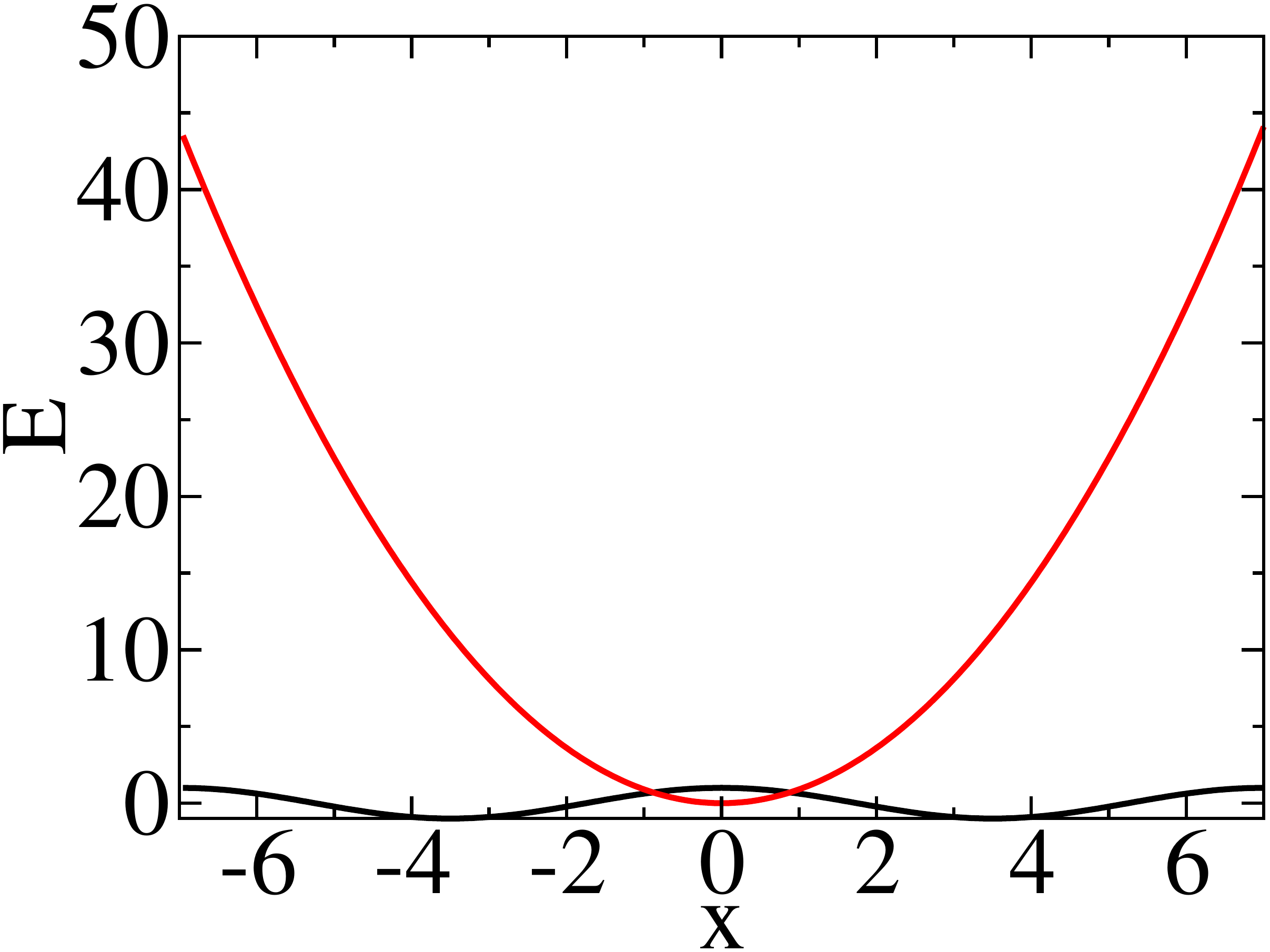}\includegraphics[scale=0.2]{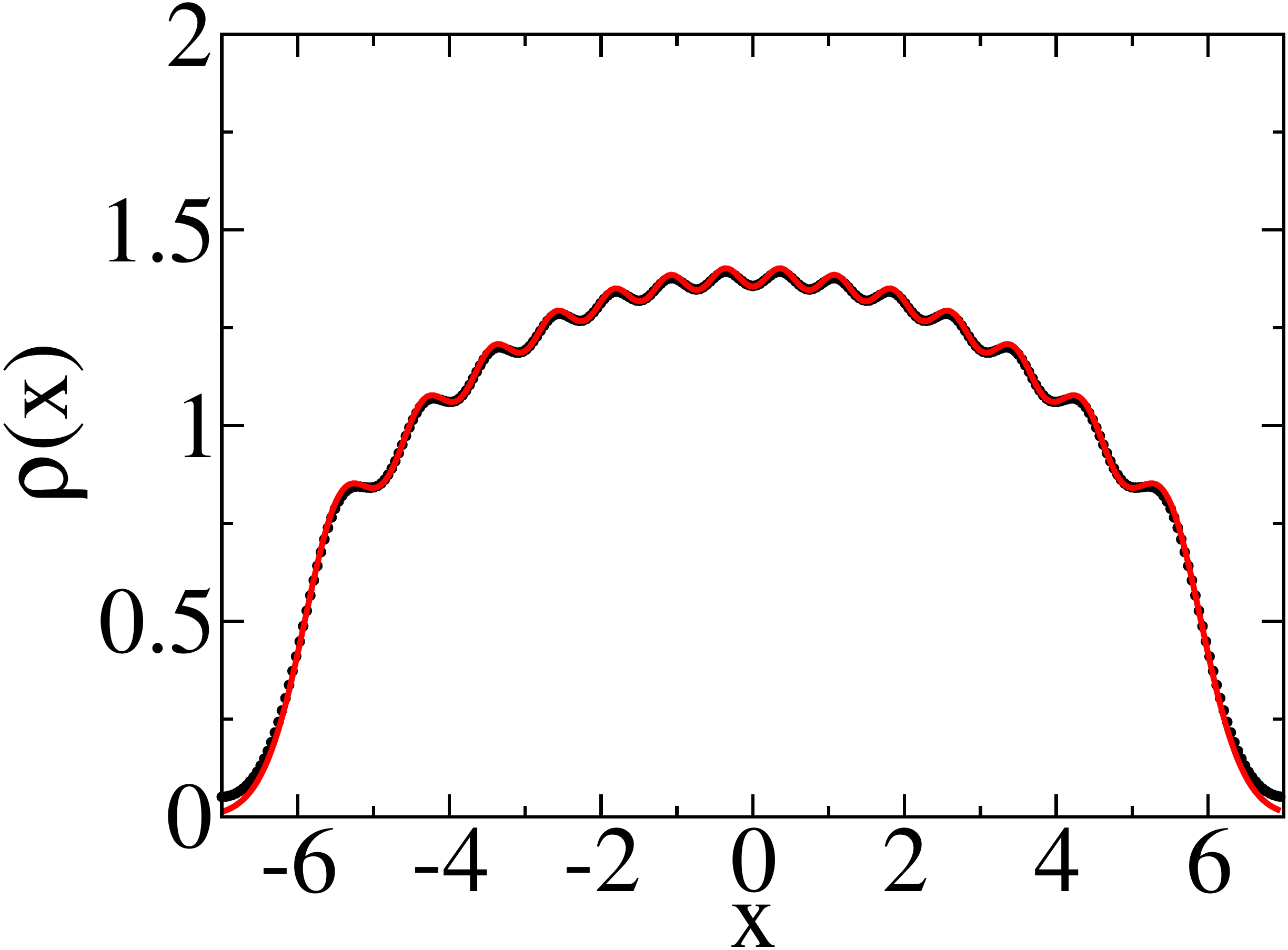}
\includegraphics[scale=0.2]{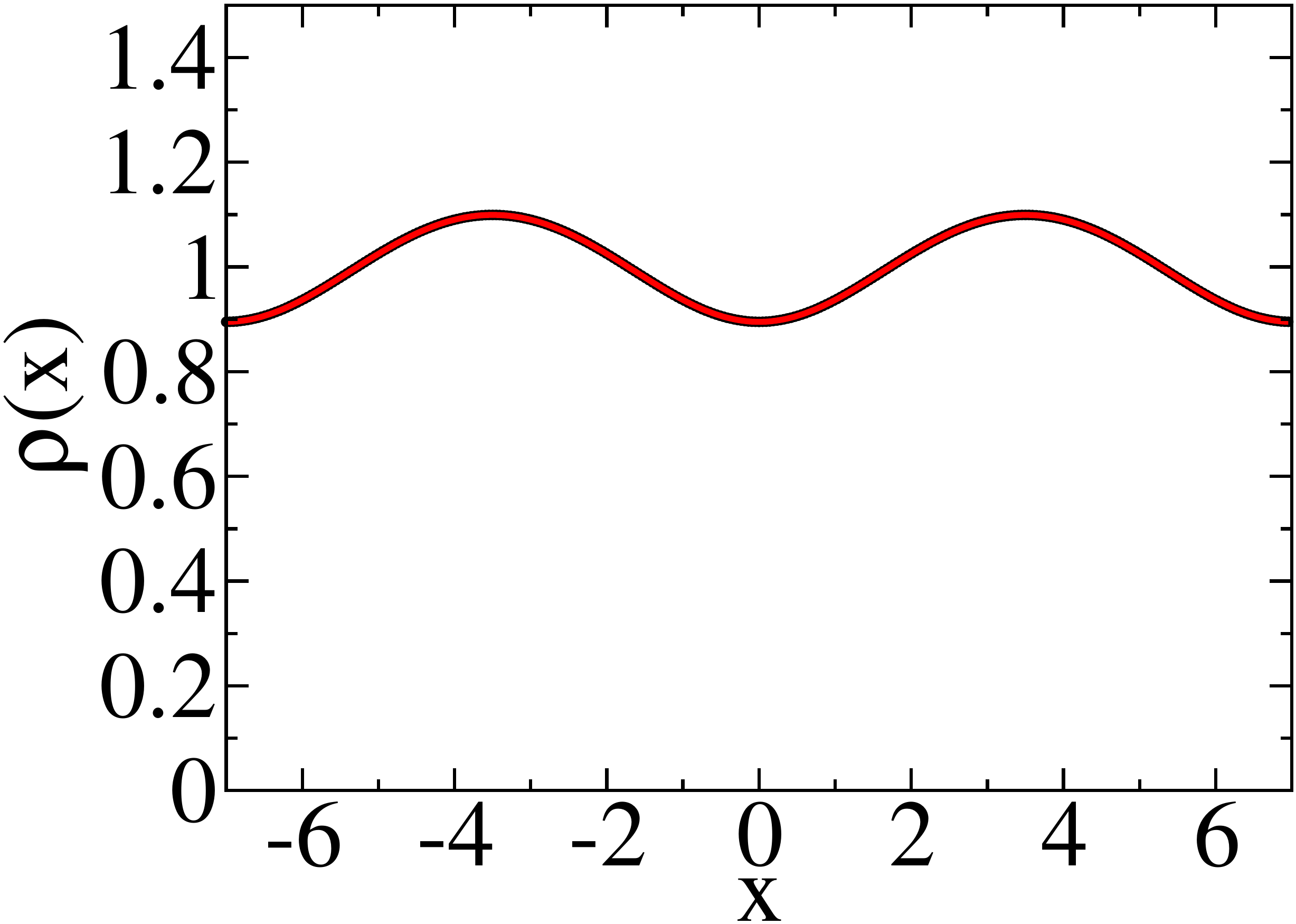}
\caption{Right: The parabolic one-body potential, $V_{\rm para}(x)=0.9 x^2$ (red), used to prepare the initial
state and the post-quench one-body cosine potential, 
$V_{\rm cos}(x)=\cos(4\pi x/L)$ with amplitude $A=1$ (black). Center: The density, $\rho(x)$, of the gas ($N=L=14,c=7200$) 
in its ground state in the presence of the parabolic potential: red (analytics), black (NRG).
Right: The density, $\rho(x)$, of the gas ($N=L=14,c=7200$) in its ground state in the presence of
the cosine potential with amplitude $A=2$: red (analytics), black (NRG).}
\label{potandgs}
\end{figure}

In this section we demonstrate that we can compute the equilibrium properties of the gas in the one-body
potentials necessary to describe the dynamics of the system post-quench.
We first show that we can prepare the initial state, the ground state of the gas in the presence
of the parabolic potential, $V_{\rm para}$, accurately.  To this end we show in the central panel of 
Fig. 2 the density profile, $\rho(x)$, of the gas in this trap.  We see that we get good agreement
between the NRG numerics for $N=L=14, c=7200$ and an analytical computation of the density profile of the gas in its Tonks-Girardeau
limit ($c=\infty)$.  Such analytics are possible because the gas, for certain quantities such as $\rho(x)$,
can be treated as equivalent to free fermions.  In Fig. 2 (right panel) we also demonstrate that we can accurately
compute the density profile of the gas in a cosine potential.

However in order to compute post-quench dynamics, we need to be able to describe not only the
ground state in the cosine potential, but some large number of excited states.
In our quench protocol, we take as our initial $t=0$ state the ground 
state of the gas in the parabolic potential, $|\psi_{GS,{\rm para}}\rangle$.  If we can compute a wide range
of eigenstates in the cosine potential, both ground and excited states, $|\psi_{\alpha,{\rm cos}}\rangle$,
we can expand this initial state in terms of the post-quench basis:
\begin{equation}
|\psi_{GS,{\rm para}}\rangle = \sum_\alpha c_\alpha |\psi_{\alpha,{\rm cos}}\rangle.
\end{equation}
Of course for this expansion to be exact, we would need to know {\it all} of the eigenstates of the
gas in the parabolic potential.  We will instead settle for a determination of the post-quench eigenbasis
that allows us to include enough states so that $\sum_\alpha |c_\alpha|^2 > 0.99$.

Because we use the eigenstates of the Lieb-Liniger model absent a one-body potential, $|\psi_{\alpha,LL}\rangle$,
as the computational basis of the NRG, the NRG gives any eigenstate in a one-body potential as a linear
combination of such states:
\begin{equation}\label{LLexpand}
|\psi_{{\rm one-body}}\rangle = \sum_\alpha b_\alpha |\psi_{\alpha,{\rm LL}}\rangle.
\end{equation}
Each Lieb-Liniger state $|\psi\rangle_{\alpha,LL}$ is characterized by $N$-rapidities, $\lambda_i,~i=1,\ldots,N$.
These rapidities govern the energy, $E_\alpha$, and momentum, $P_\alpha$, of the state (relative to the Hamiltonian, $H_{LL}$):
\begin{equation}
E_\alpha = \sum^N_{i=1} \lambda_{\alpha,i}^2; ~~~ P_\alpha = \sum^N_{i=1} \lambda_{\alpha,i}.
\end{equation}
These rapidities are found as solutions of the Bethe equations:
\begin{equation}\label{Beqn}
e^{i\lambda_nL} = \prod_{m\neq n}^{N} \frac{\lambda_n-\lambda_m+ic}{\lambda_n-\lambda_m-ic},~~n=1,\ldots,N.
\end{equation}
In the limit of $c=0$, we see that the Bethe equations collapse to the momentum quantization condition for
a particle in a periodic system of length L.

\begin{figure}\label{spectrum}
\centering
\includegraphics[scale=0.4,angle=0]{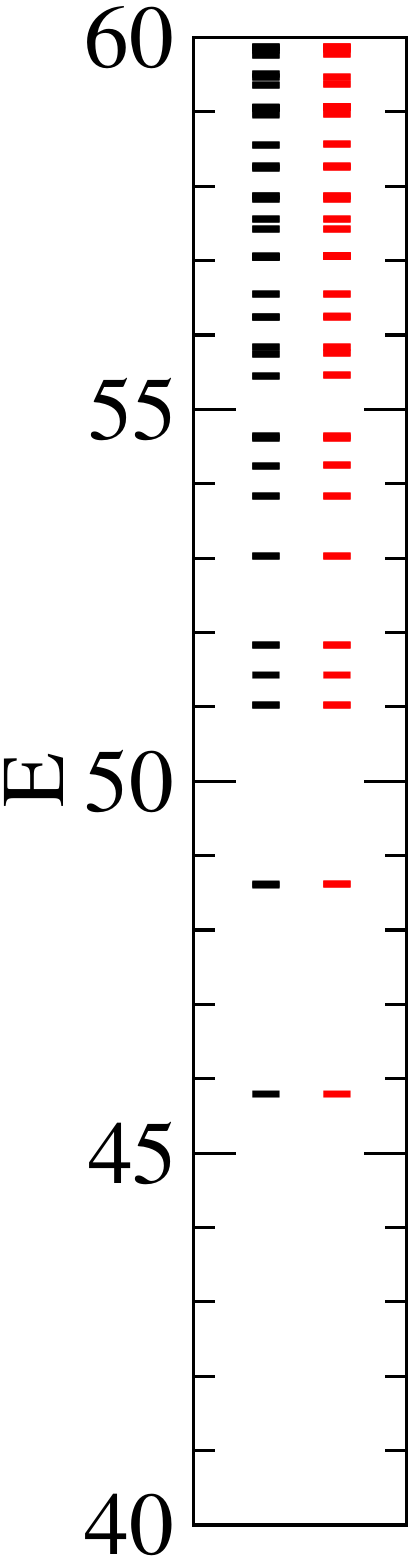}
\caption{A plot of the energies of the gas $(N=L=14, c=7200)$ in the cosine potential as determined
by the NRG (black) and by analytics (red) in the hardcore limit $(c=\infty)$.}
\end{figure}

In computing the spectrum of states in the cosine potential, we employ the variant of the NRG discussed
in Ref. \cite{NRG2}.  The NRG in its plain vanilla formulation \cite{NRG} 
can compute the spectrum of the low lying states of the gas in the
one-body potential \cite{NRG4}.  But to capture accurately an appreciable fraction of the spectrum, we need
to employ a sweeping routine \cite{NRG2} analogous to that used in the finite volume routine of the density
matrix renormalization group \cite{DMRG}.

In Fig. 3 we present the results for the spectrum of the gas in the $V_{\rm cos}(x)=\cos (4\pi x/L)$ ($A=1$) as computed 
with the NRG and with analytics in the $c=\infty$ limit.  We see
we are able to describe accurately a wide range of the spectrum.  For higher energy states, there are some
slight differences between analytics and the NRG which we believe can be ascribed to $1/c$ corrections, which while
small are still present.

\section{Post-quench dynamics of conserved charges}

\begin{figure}[t]
\centering
\includegraphics[scale=0.6]{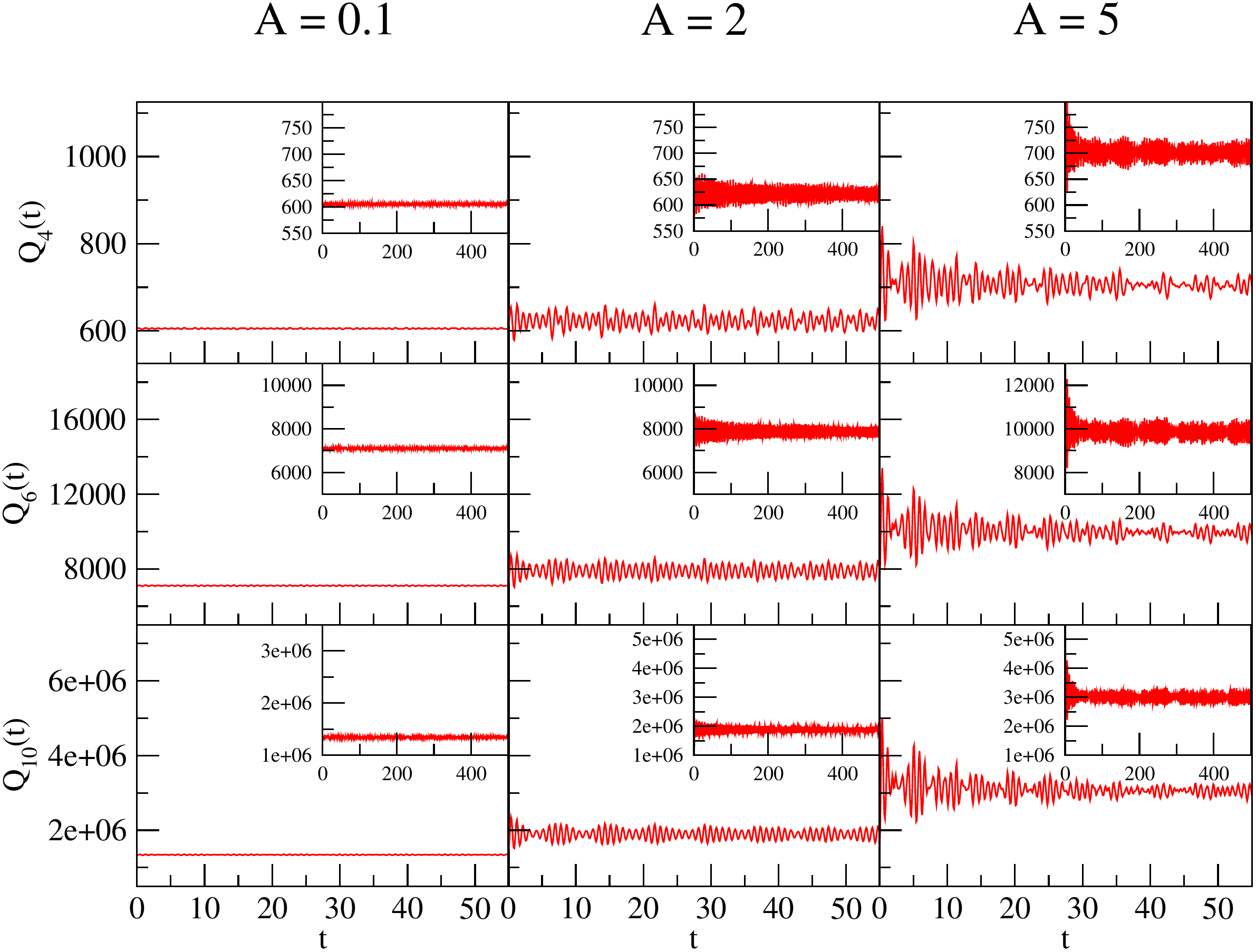}
\caption{The time evolution of the conserved charges, $Q_n$, of the Lieb-Liniger model post-quench.  Shown
are $Q_4$, $Q_6$, and $Q_{10}$ for a release into a cosine potential of amplitude $A=0.1,2,$ and $5$.
The frequency of the cosine potential, $\omega$, is set to $4\pi/L$.}
\end{figure}
In this section we consider the time evolution of the Lieb-Liniger conserved quantities, $Q_n$.  These quantities
commute with the Lieb-Liniger Hamiltonian, $[Q_n,H_{LL}]$, but in the presence of the one-body cosine potential
they become time dependent.  
\begin{figure}[t]
\centering
\includegraphics[scale=0.6]{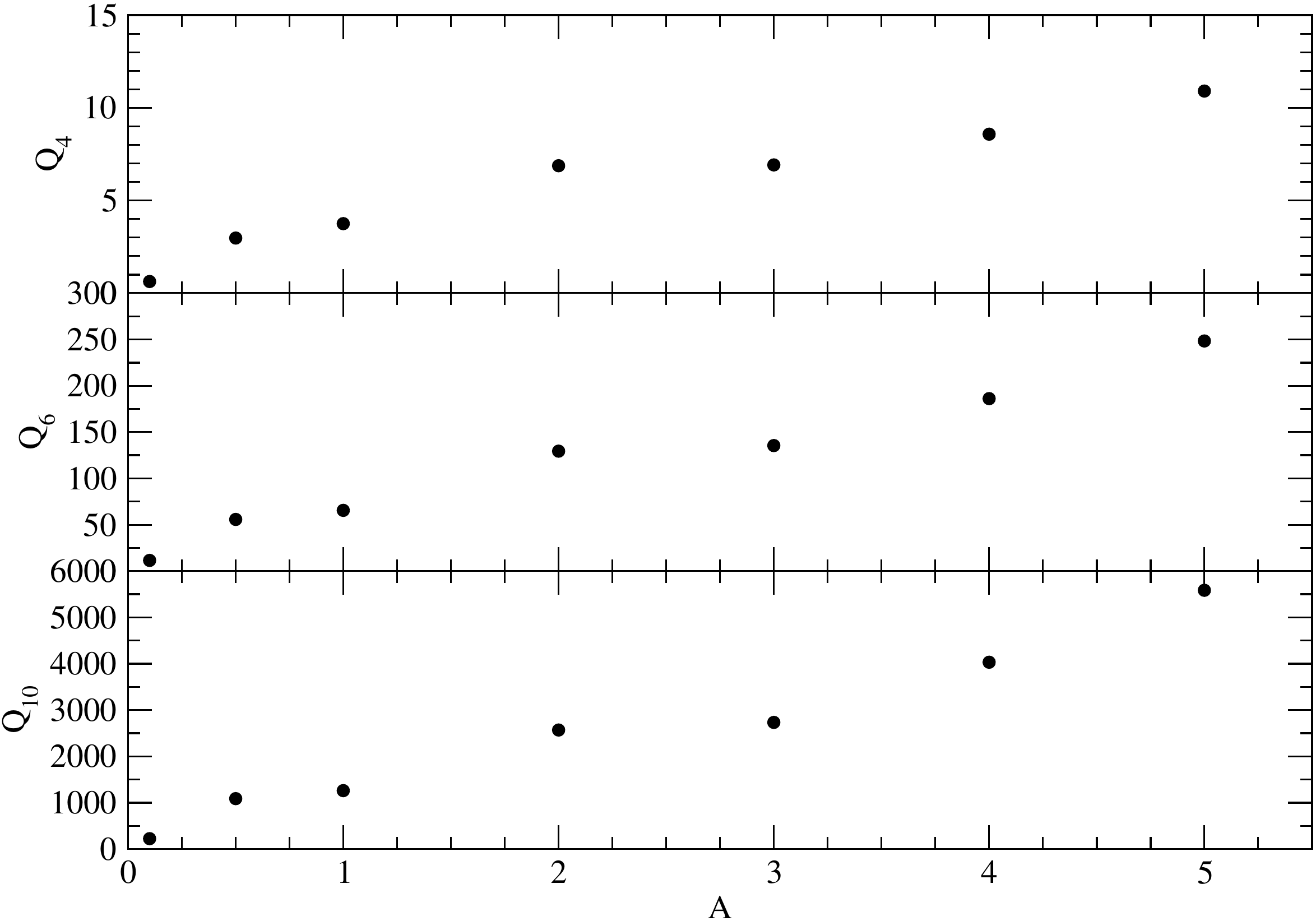}
\caption{Amplitudes of the oscillations of the conserved charges, $Q_4$, $Q_6$, and $Q_{10}$, as a function
of the amplitude, $A$, of the post-quench cosine potential.}
\end{figure}
To compute this time evolution, we first note the time evolution of our initial $t=0$ state is expressible
as
\begin{equation}
|\psi_{GS,{\rm para}}(t)\rangle = \sum_\alpha c_\alpha |\psi_{\alpha,{\rm cos}}\rangle e^{itE_{\alpha,{\rm cos}}}.
\end{equation}
To compute 
$$
Q_n(t) = \langle \psi_{GS,{\rm para}}(t)|Q_n|\psi_{GS,{\rm para}}(t)\rangle,
$$ 
we then need to know
$\langle \psi_{\beta,{\rm cos}}|Q_n|\psi_{\alpha,{\rm cos}}\rangle$.  But because $|\psi_{\alpha,\rm{cos}}\rangle$
is given in terms of Lieb-Liniger states (Eqn. \ref{LLexpand}), this amounts to knowing the action
of the charges, $Q_n$, on such states.
This however is straightforward \cite{korepin}:
\begin{equation}
Q_n|\psi_{LL,\alpha}\rangle = \sum^N_{i=1}\lambda^n_{\alpha,i}|\psi_{LL,\alpha}\rangle.
\end{equation}
The first two charges in the sequence, $Q_1$ and $Q_2$, give the momentum and energy respectively of
the Lieb-Liniger state.  Because of the system's parity symmetry, $x \rightarrow -x$, all of the odd charges
evaluate to zero on $|\psi_{LL,\alpha}\rangle$.  We thus will focus on the even charges, $Q_{2n}$, alone.

\begin{figure}[t]
\centering
\includegraphics[scale=0.6]{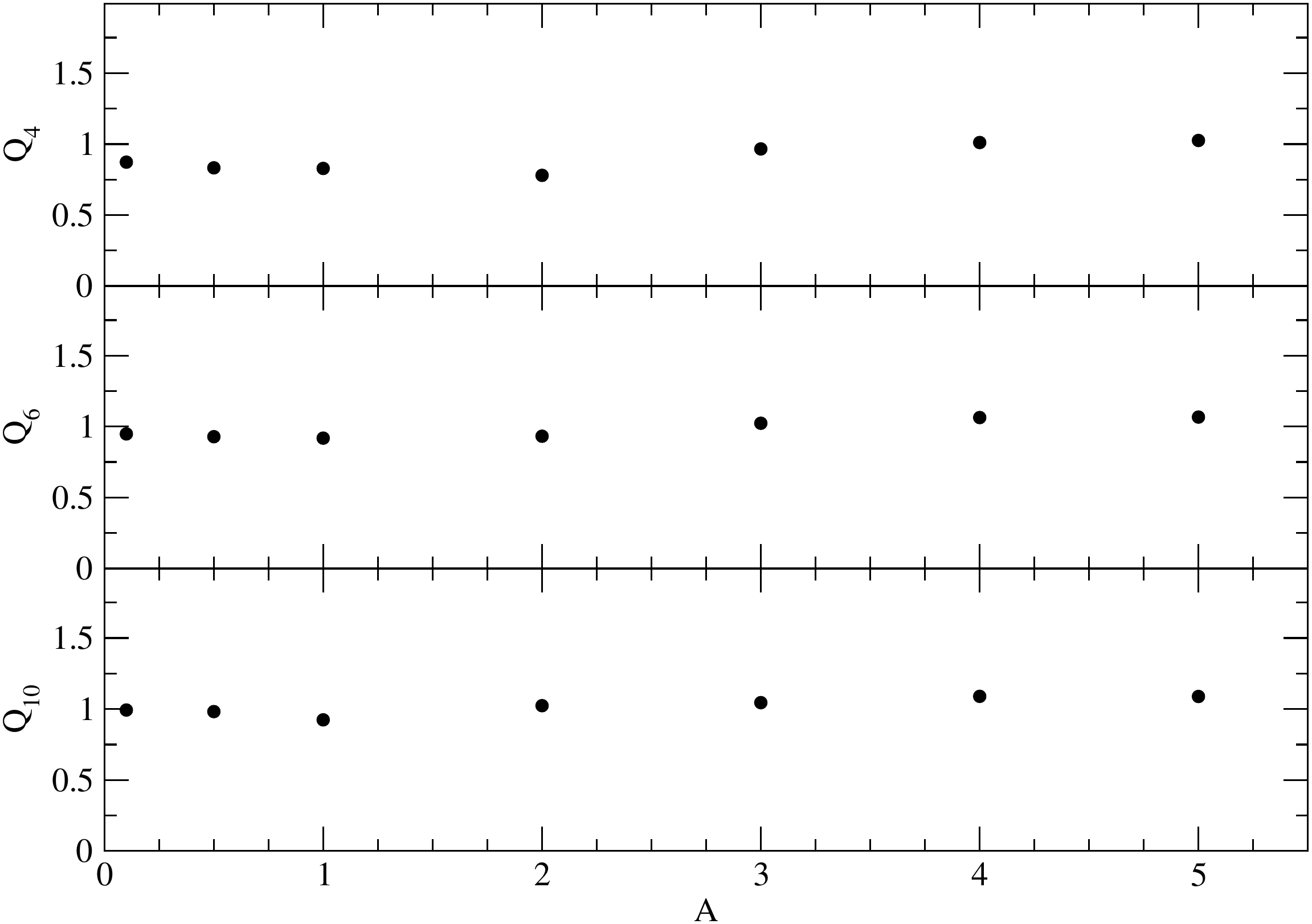}
\caption{Average frequency of the oscillations of the conserved charges, $Q_4$, $Q_6$, and $Q_{10}$ as a function
of the amplitude, $A$, of the post-quench cosine potential.}
\end{figure}

In Fig. 4 we present the time evolution of three charges $Q_4$, $Q_6$, and $Q_{10}$ for quenching into cosine
potentials with amplitudes $A=0.1,2,$ and $5$.  We see the charges oscillate in time with increasing amplitude 
as the amplitude of the integrability breaking cosine potential is increased.  These oscillations occur
about a well defined mean.  This mean smoothly increases from its $A=0$ value as we increase the strength of the cosine
potential.  We also see for small times, the expectation values of the charges
are characterized by transients, but then settle into a steady state oscillatory behavior.

In Fig. 5 we plot the amplitudes of the oscillations of the charges as a function of the amplitude of the cosine potential.
We see that these amplitudes of the oscillations are (roughly) linearly related to the amplitudes, $A$'s, of the cosine.  In the limit
that $A$ vanishes, the charges become constants of motion, as expected.  More interestingly, however, the frequencies at which the
charges are oscillating, Fig. 6, are independent of the amplitude of the cosine potential.  Instead these frequencies
are roughly but uniformly equal to that of the cosine
potential itself, $4\pi/L \sim 1$.

\section{Discussion and future directions}

In this brief report we have considered post-quench dynamics of a cold atomic gas quenched from its ground state in a parabolic trap
to a cosine shaped trap.  To characterize these dynamics we have investigated the time dependency of the Lieb-Liniger conserved quantities
induced by the presence of the post-quench integrability breaking cosine.  In the presence of integrability breaking, the charges oscillate
about a mean.  This mean behaves smoothly as a function of the strength of integrability breaking, the amplitude, $A$, of the cosine.  
Similarly the amplitude of the oscillations go smoothly to zero as $A$ is reduced to zero.  However for the frequency of the oscillations
of the charges this is not the case.  Instead this frequency is related directly to the cosine potential's own frequency, which in our
quench protocol is kept constant.  

Our findings for the behavior of the charges are then in accordance with Ref. \cite{yur_ols,kollar,canovi}, namely we find that the introduction
of integrability breaking leads to a smooth interpolation of the expectations values of observables (here the charges) 
between that in the integrable limit and that in the fully chaotic limit where the standard thermodynamic ensembles govern dynamics.  However we also
find that the time scale, $\tau_{IB}$, for integrability breaking is not necessarily associated with the strength of the integrability breaking.
We find rather that the charges oscillate with a frequency independent of the amplitude of the one-body cosine potential.
This suggests the intriguing possibility that even relatively large integrability breaking terms, if suitably low-frequency, might
lead to long thermalization times before completely chaotic behavior is observed.

In future work we intend to explore these questions in terms of the momentum distribution function (MDF) of the gas.
The long time behavior of the MDF is considerably more complicated to compute than the time dependency of the charges.  In the presence
of integrability breaking, it requires one to compute a large number of matrix elements of the form,
$$
\langle \alpha, {\rm LL}|\psi (x)\psi^\dagger(0)|\beta,{\rm LL}\rangle,
$$
where $\psi(x)$ is the Bose field operator.
Each of these matrix elements is evaluated by inserting a resolution of the identity between the fields $\psi$, itself a computationally
intensive task \cite{cc}.  Having already carried out preliminary computations for systems sizes
of $N=L=14$, we nonetheless expect to be able to perform these calculations for system sizes of up to $N \sim 25-30$.

\acknowledgments
G.P.B. and J.-S.C. acknowledge support from the Netherlands Organisation for Scientific Research (NWO). 
R.M.K. acknowledges support by the US DOE under contract DE-AC02-98CH10886.

\end{document}